\documentclass[epj,twocolumn]{webofc}
\usepackage[varg]{txfonts}   
\woctitle{NSRT15}
\begin{document}
\title{Superfluid fission dynamics with microscopic approaches}
%
%

\author{C. Simenel\inst{1}\fnsep\thanks{\email{cedric.simenel@anu.edu.au}} \and
        G. Scamps \inst{2,3}
             \and
        D. Lacroix\inst{4}
             \and
        A. S. Umar\inst{5}
}

\institute{Department of Nuclear Physics, Research School of Physics and Engineering \\ Australian National University, Canberra, Australian Capital Territory 2601, Australia
\and
           GANIL, CEA/DSM and CNRS/IN2P3, Bo\^ite Postale 55027, 14076 Caen Cedex, France
\and
           Department of Physics, Tohoku University, Sendai 980-8578, Japan
\and
	Institut de Physique Nucl\'eaire, IN2P3-CNRS, Universit\'e Paris-Sud, F-91406 Orsay Cedex, France
\and
	Department of Physics and Astronomy, Vanderbilt University, Nashville, Tennessee 37235, USA
          }

\abstract{%
Recent progresses in the description of the latter stage of nuclear fission are reported. 
Dynamical effects during the descent of the potential towards scission and in the formation of the fission fragments are studied with the time-dependent Hartree-Fock approach with dynamical pairing correlations at the BCS level.
In particular, this approach is used to compute the final kinetic energy of the fission fragments. 
Comparison with experimental data on the fission of  $^{258}$Fm  are made. }
\maketitle
\section{Introduction}
\label{intro}

Despite important progresses  since the discovery of nuclear fission in 1939  \cite{hah39,mei39}, it remains an important challenge for theorists.  
Compared with other nuclear dynamical processes such as fusion, the overall time-scale for  fission is relatively long.
This  suggests that the evolution across the potential energy surface is a slow process.
However, it is also expected that the systems could encounter rapid shape evolution  in the later stages near scission. 
Theoretical prediction of the fission fragments and their characteristics are often based on the adiabatic approximation \cite{mor91}. 
It is assumed that the internal degrees of freedom are equilibrated while the system evolves along the fission path.  
However, the shape evolution near scission is expected to be non-adiabatic and the approximation may break down in the latter stage of fission \cite{riz13,dub08}. 
Thus, the inclusion of the dynamical effects near scission are  crucial.
In particular, the dynamics has been shown to generate most of the excitation energy in the fragments \cite{sim14}.

Microscopic approaches are well suited to investigate the dynamics of nuclear fission \cite{gou05,bon06,dub08,sta09,pei09,you11,war12,abu12,mir12,sta13,mcd13,sad13,mcd14,sim14,sch14,sch15,tan15}. 
Microscopic nuclear dynamics is usually treated at the mean-field level with the time-dependent Hartree-Fock theory (see \cite{neg82,sim12b} for reviews). 
In particular, Negele {\it et al.}, have demonstrated the possibility to account for important dynamical effects in the fission process within the time dependent Hartree-Fock approximation \cite{neg78}. 
Similar approaches have recently been used by several groups to study fission \cite{sim14,sca15,god15} and quasi-fission \cite{sim12a,wak14,obe14,uma15,was15,ham15}.
Static or dynamical pairing correlations are sometimes included in mean-field calculations \cite{ave08,eba10,ste11,has12,sca13}. 
Here, the superfluid dynamics is obtained using a recent code solving the time-dependent Hartree-Fock equation with BCS dynamical pairing correlations \cite{sca13}. 

The importance of the dynamical effects  is investigated in fission of  $^{258,264}$Fm \cite{sim14,sca15}. 
It is shown that dynamics has an important effect on the scission configuration and on the kinetic and excitation energies of the fragments. 
The vibrational modes of the fragments in the post-scission evolution are also analyzed.
Quantum shell effects are shown to play a crucial role in the dynamics and formation of the fragments. 

\section{Adiabatic calculations}

The adiabatic evolution is described in the traditional way by minimizing the HF energy under an external constraint inducing elongation of the system up to the scission point. This constraint can be on the distance $R$ between the centres of mass of the fragments (defined assuming a sharp cut at the neck), or on multipole moments. 
The static Hartree-Fock equation is solved with BCS pairing correlations  using the \textsc{ev8} code~\cite{bon05}.
The SLy4$d$ parametrization  \cite{kim97} of the Skyrme functional \cite{sky56} is used with a surface pairing interaction \cite{bender03}. 
The calculations are performed on a Cartesian grid with mesh size $0.8$~fm. 

\begin{figure}
\centering
\includegraphics[width=8cm]{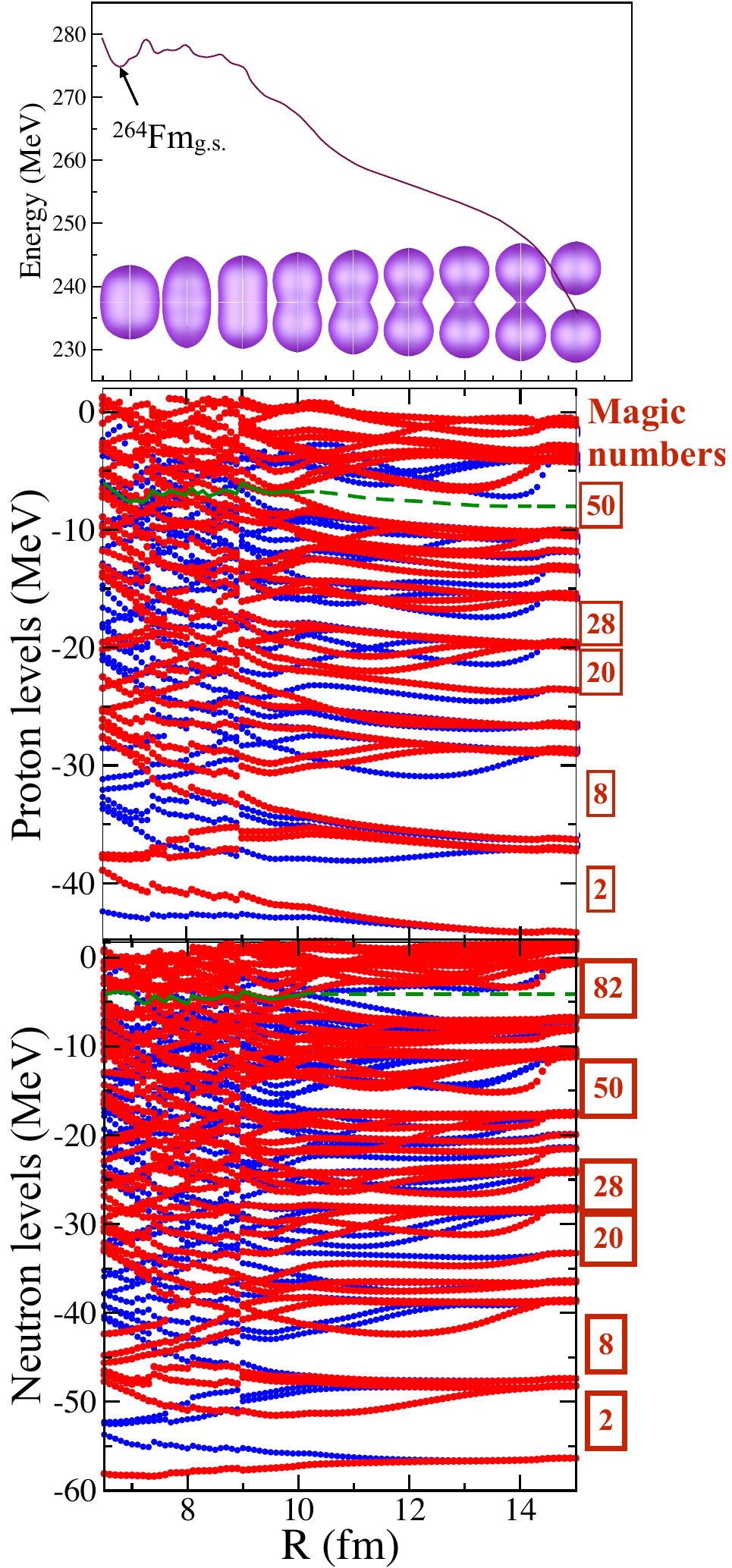}
\caption{Adiabatic calculation of $^{264}$Fm symmetric fission with the SLy4$d$ Skyrme functional and surface pairing. The distance between the fragments $R$ is defined in Ref.~\cite{sim14}. (top) Adiabatic potential and isodensities at half the saturation density $\rho_0/2=0.08$~fm$^{-3}$. The energy is defined with respect to the asymptotic final $^{132}$Sn+$^{132}$Sn state. Single-particle levels are plotted for protons and neutrons in the middle and bottom panels, respectively. Positive and negative parity states are shown in red and blue, respectively. Adapted from \cite{sim14}.}
\label{fig-1}       
\end{figure}

\begin{figure}
\centering
\includegraphics[width=8cm]{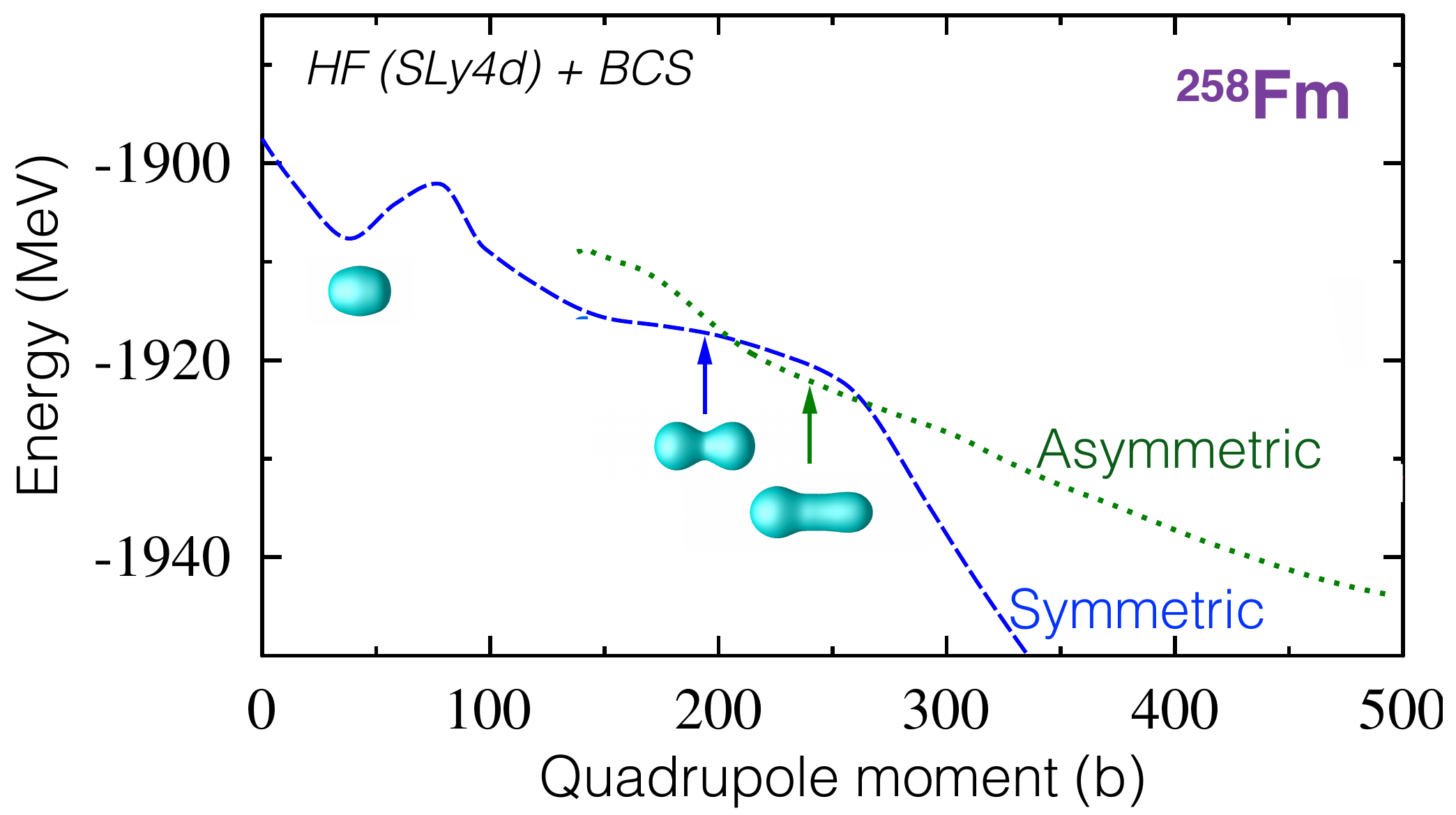}
\caption{Adiabatic potential in $^{258}$Fm  fission with the SLy4$d$ Skyrme functional and surface pairing as a function of the total quadrupole moment calculated along the fission axis. Two fission valleys are represented: one leading to symmetric fragments (dashed line) and one to asymmetric fragments (dotted line). Adapted from \cite{sca15}.}
\label{fig-2}       
\end{figure}

The   adiabatic potential has been calculated in Ref.~\cite{sim14} for $^{264}$Fm and in Ref.~\cite{sca15} for $^{258}$Fm.
The results are reported on Figs.~\ref{fig-1} and \ref{fig-2} for $^{264}$Fm and $^{258}$Fm, respectively. 
In both cases, the resulting barrier is in good agreement with other calculations \cite{mol09,pei09,sta09,kow10,sad13}. 

An important question is which configuration should we choose as initial state of the time-dependent calculations. 
A possible choice would be to start with a configuration where the fragments have established their identity. 
This may be difficult to determine from the total density. 
However, it is possible to see where fragments are pre-formed  from the single-particle levels \cite{sim14}. 
In the case of $^{264}$Fm, the proton and neutron single-particle energies are plotted in the middle and bottom panels of Fig.~\ref{fig-1}, respectively. 
Many levels cross at short distance ($R<10$~Fm). 
We also observe sudden jumps in the single-particle energies due to changes of macroscopic shapes. 
At larger distances, we observe the appearance of energy gaps associated with the spherical magic numbers.
As the $^{132}$Sn fragments are doubly magic, there is no crossing of the Fermi surface after the magic gaps at $Z=50$ and $N=82$ are both formed. 
This occurs at $R\sim10.5$~fm. 

\begin{figure}
\centering
\includegraphics[width=8cm]{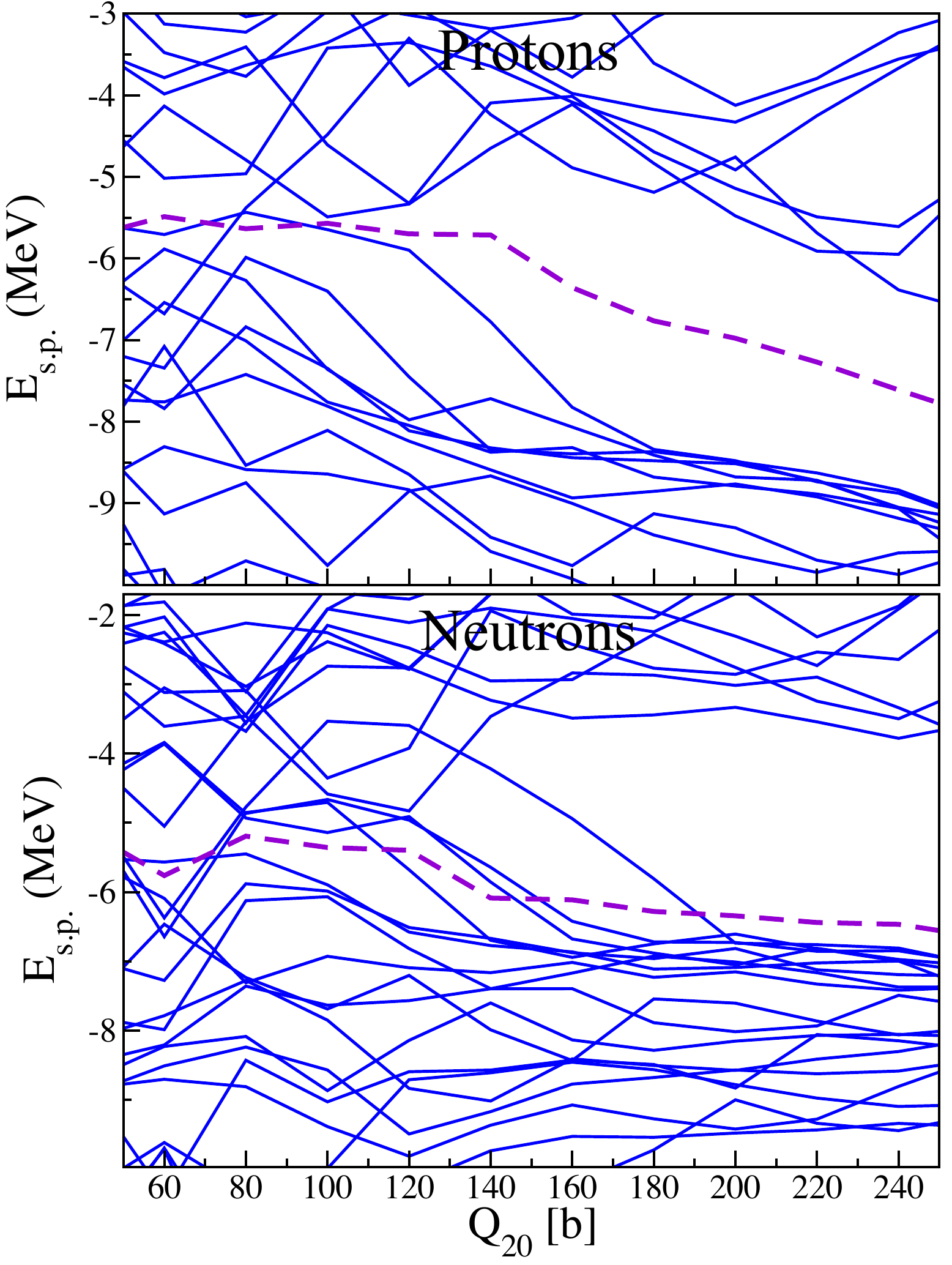}
\caption{Adiabatic calculation of $^{258}$Fm symmetric fission with the SLy4$d$ Skyrme functional and surface pairing. Single-particle levels are plotted for protons and neutrons in the top and bottom panels, respectively. The dashed line shows the Fermi level.}
\label{fig-3}       
\end{figure}

\begin{figure}
\centering
\includegraphics[width=8cm]{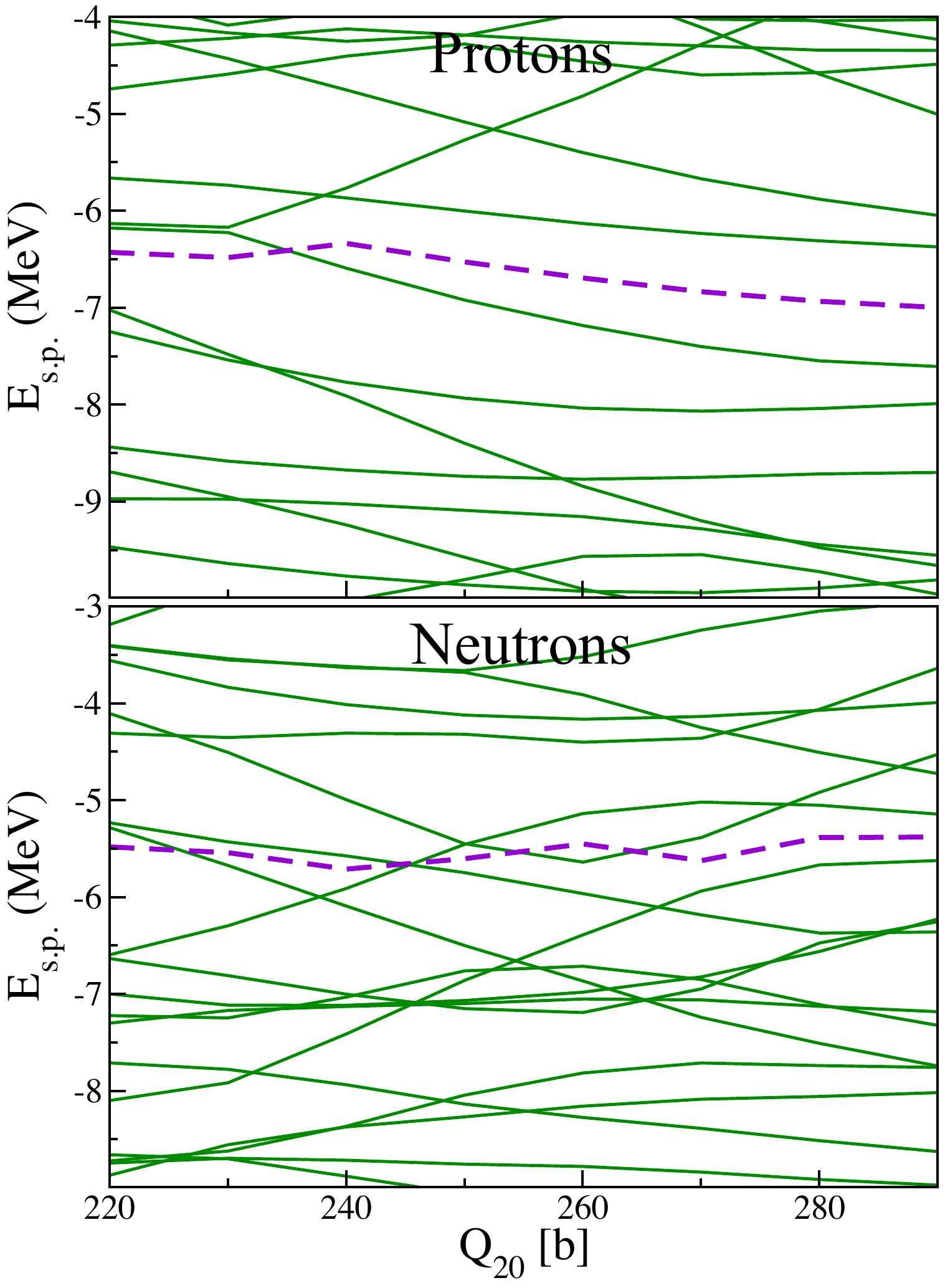}
\caption{Same as Fig.~\ref{fig-4} for the asymmetric fission valley of $^{258}$Fm.}
\label{fig-4}       
\end{figure}

The symmetric fission mode of $^{258}$Fm   is also driven by spherical shell effects \cite{hul89}.
Indeed, the fragments  have a proton magic number $Z=50$. 
The evolution of proton and neutron single-particle levels are shown in Fig.~\ref{fig-3} for the symmetric mode.
As in the $^{264}$Fm case, we observe the formation of $Z=50$ and $N=82$ magic gaps after some global elongation of the fissioning system (quantified by the quadrupole moment) has been reached. 
However, in the case of neutrons, the levels near the Fermi surface (represented with a dashed line) are only partially filled. 

The case of $^{258}$Fm    asymmetric fission is more complicated as the fragments are not magic and therefore can be deformed. 
Indeed, the associated single-particle energies plotted in Fig.~\ref{fig-4} show no energy gap neither for protons nor for neutrons. 
Therefore, it is not always straightforward to identify the pre-formation of the fragments simply by looking at the formation of spherical magic gaps.

\section{TDHF+BCS calculations}

In general, it is necessary to consider a range of initial conditions in order to investigate the dynamical evolution along the fission path with time-dependent calculations. 
Starting with a more compact configuration is desirable in order to capture most of the dynamical effects in the evolution along the fission path. 
However, the TDHF+BCS approach accounts only for part of the correlations and it is likely that some correlations will be missing for the most compact configurations. 
The computational time is also a practical limitation.
However, starting with a configuration too close to scission, we would miss some of the dynamical effects.
In particular, the final observables (mass and charge distributions, total kinetic energy...) may exhibit a dependence with the initial configuration when the latter is too close to scission \cite{sca15,god15}.

A compromise has to be found by investigating several initial conditions. 
As an example, we found that the final properties of the fragments in $^{258}$Fm asymmetric fission were essentially independent on the initial condition if the latter was chosen to be in the range $210$~b$<Q_{20}<270$~b (see Fig.~\ref{fig-2}).
The sensitivity to the initial condition (in TDHF calculations without dynamical pairing) has also recently  been investigated in details by Goddard and collaborators in ref.~\cite{god15}.

\begin{figure}
\centering
\includegraphics[width=8cm]{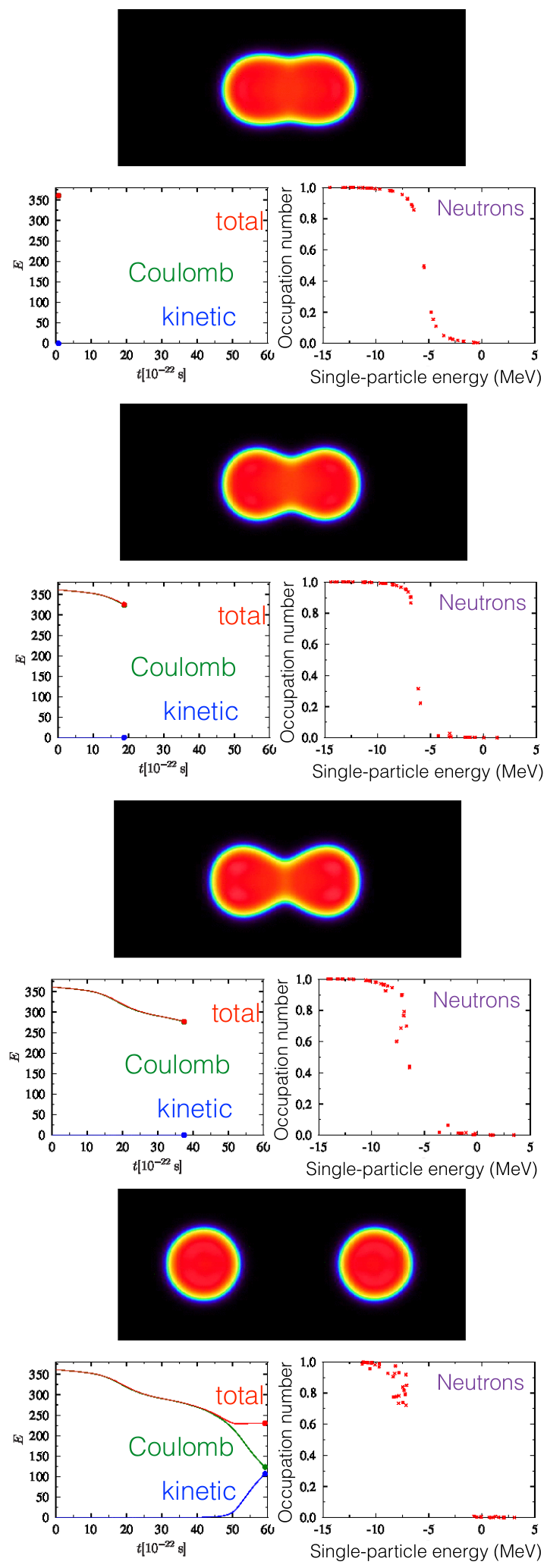}
\caption{TDHF with dynamical BCS pairing calculation of $^{258}$Fm symmetric fission with the SLy4$d$ Skyrme functional and surface pairing interaction. The density profile, energy repartition of the fragments and single-particle neutron occupation number distributions are shown at four different times. }
\label{fig-5}       
\end{figure}

The importance of dynamical pairing correlations in the evolution was demonstrated in Ref.~\cite{sca15} at the BCS level.
The time-dependent BCS equations, giving the evolution of the single-particle occupation numbers, were solved simultaneously with the TDHF equation for the evolution of the single-particle wave-functions. 
An example of the resulting dynamics of the occupation numbers is shown in Fig.~\ref{fig-5} for the symmetric fission mode of $^{258}$Fm.
An important rearrangement of the neutron occupation numbers is observed during the evolution. 
As a result, the time variation of the occupation numbers induces an evolution of the pairing energy \cite{sca15}.

\begin{figure}
\centering
\includegraphics[width=7cm]{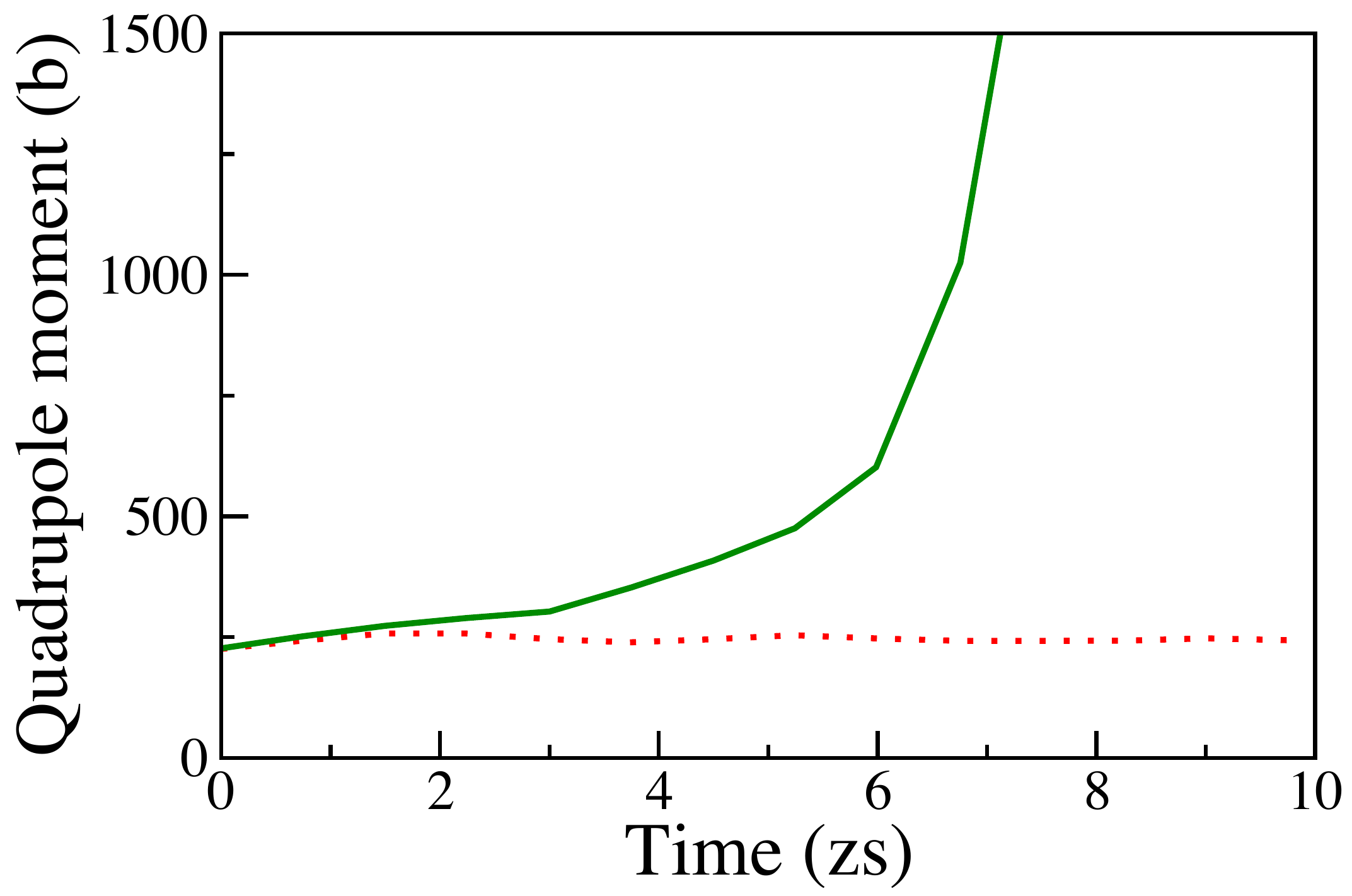}
\caption{Time-dependent Hartree-Fock with dynamical BCS pairing (TDBCS) and in the frozen occupation approximation (FOA) calculations of the total quadrupole moment in $^{258}$Fm asymmetric fission with the SLy4$d$ Skyrme functional and surface pairing interaction. }
\label{fig-6}       
\end{figure}

Accounting for dynamical pairing is also crucial for fission to occur. 
This is illustrated in Fig.~\ref{fig-6} which shows the evolution of the quadrupole moment as a proxy for the total elongation of the system. 
In the frozen occupation approximation (FOA), i.e., assuming that the single-particle numbers are those of the initial configuration, the system is not always able to fission. 
In addition, trajectories leading to fission in the FOA exhibit a significant dependence of the final observables with the choice of the initial condition \cite{god15}.

An interesting feature of the dynamical calculations is the possibility to predict the total kinetic energy. 
As discussed above, the latter is essentially independent of the initial condition if it is not too close to scission and if pairing correlations are included dynamically. 
The total kinetic energy of the fragments is computed from the sum of their Coulomb repulsion and kinetic energy. 
In particular, these quantities are well defined after scission as the nuclear interaction between the fragments vanishes due to its short range nature. 
Indeed, we see in Fig.~\ref{fig-6} that the total energy is constant after scission. 
It is equal to the final total kinetic energy (TKE) as the Coulomb energy is transformed into kinetic energy at large distances. 

\begin{figure}
\centering
\includegraphics[width=6cm]{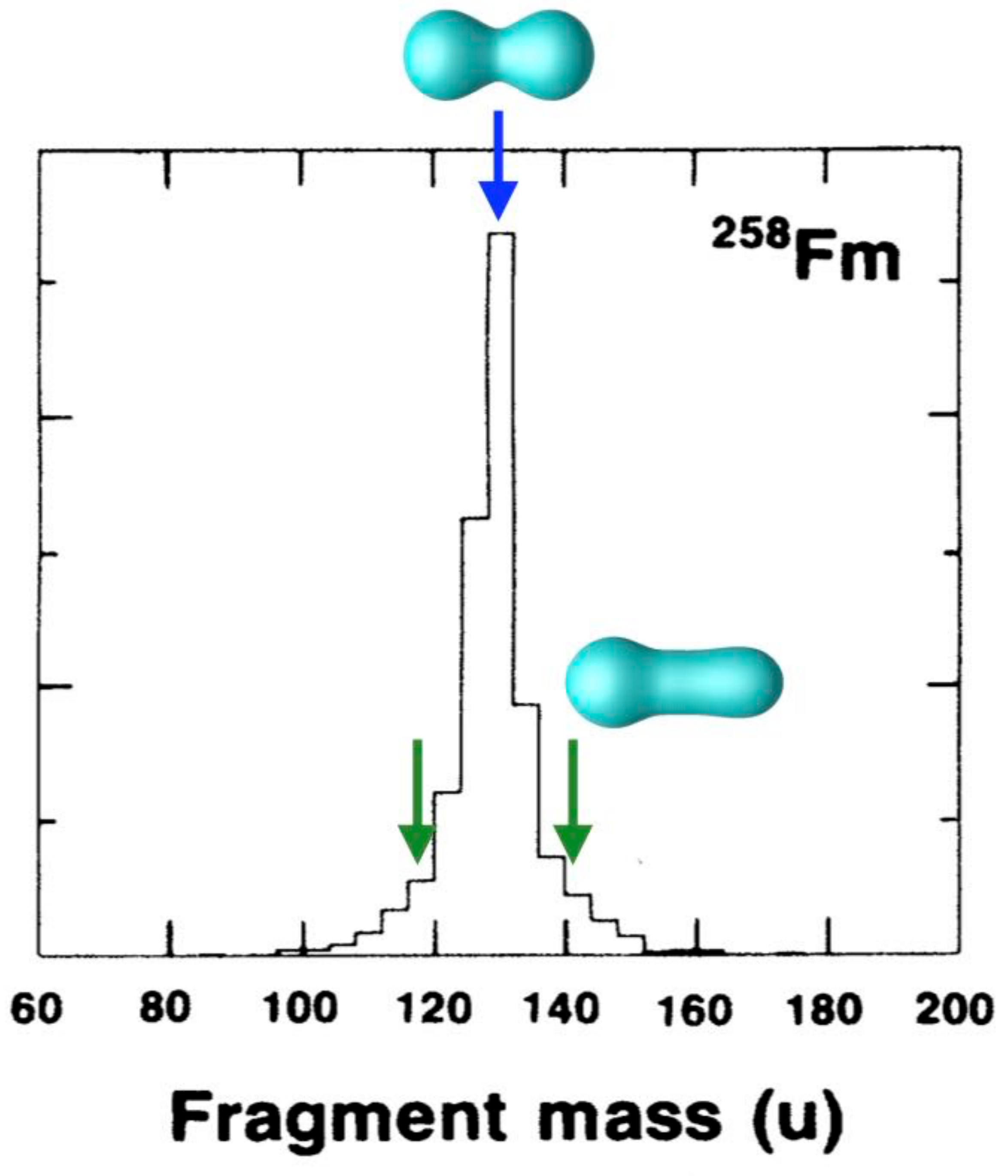}
\includegraphics[width=6cm]{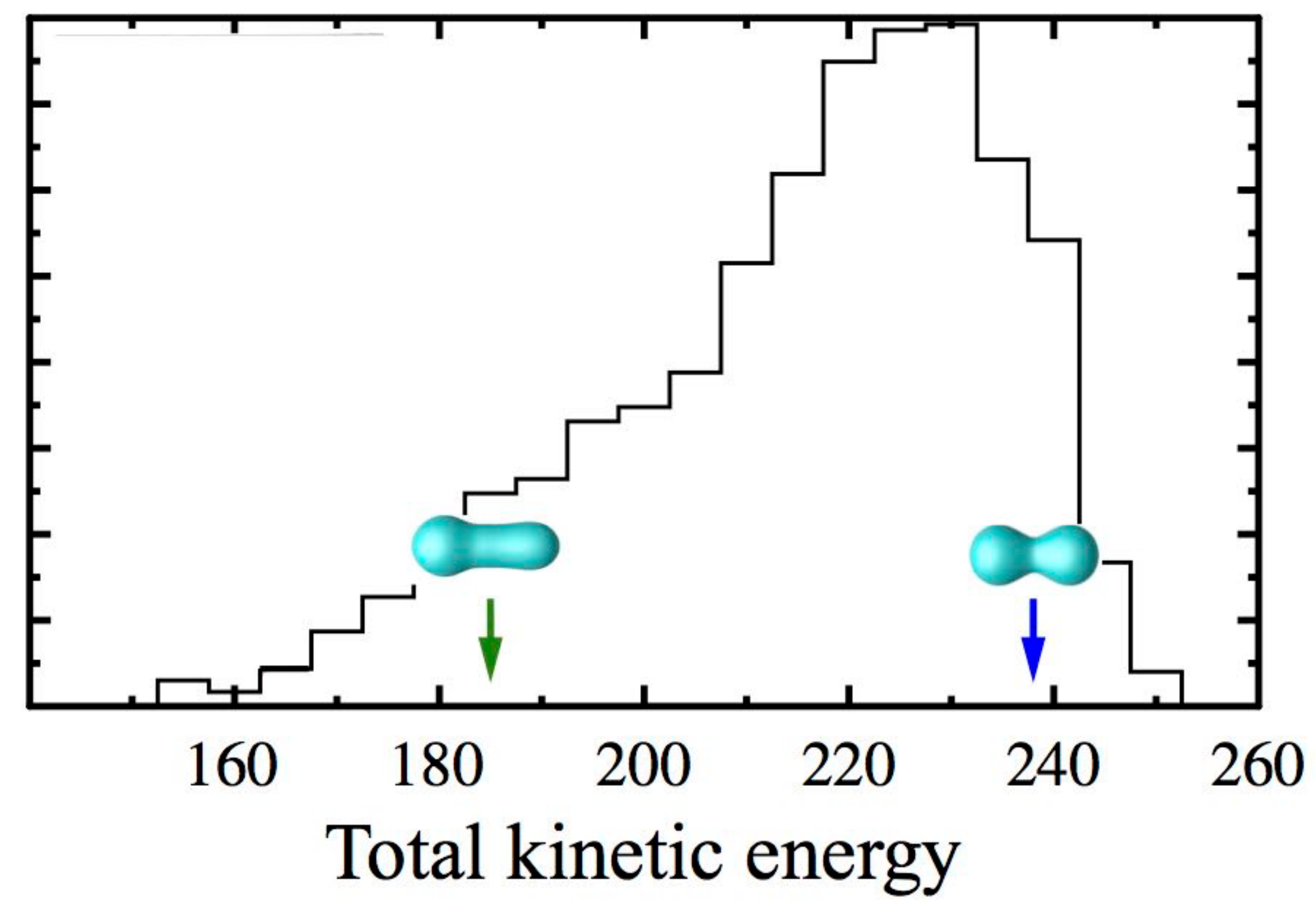}
\caption{Experimental mass (top) and TKE (bottom) distributions in $^{258}$Fm fission from Ref.~\cite{hul89}. The arrows show the TDHF predictions with dynamical pairing correlations.}
\label{fig-7}       
\end{figure}

Figure~\ref{fig-7} shows a comparison of the predicted fragment mass (top) and TKE (bottom) \cite{sca15} with experimental distributions \cite{hul89} in $^{258}$Fm fission.
The experimental data present a tail in the mass distribution interpreted as an asymmetric mode in competition with the dominant symmetric mode \cite{hul89}. 
The calculations along the  asymmetric  valley predict masses in agreement with this tail.
The calculated TKE for both modes also agree with the  two main contributions in the experimental distribution, namely a peak at high TKE and a tail at low energy. 
According to the calculations, the high TKE  mode is associated with the symmetric compact fission which is sensitive to the spherical shell effects in the $^{132}$Sn region, while the asymmetric mode leads to a much lower TKE. 
This interpretation agrees with the measured correlations between masses and TKE of the fragments \cite{hul89}.

Finally, the time-dependent calculations can also be used to investigate what forms of excitation energy are present in the fragments. 
The excitation energy is expected to be shared between deformation, collective vibration and rotation, and non-collective excitations \cite{sch11}. 
The TDHF approach has been widely used to study giant resonances \cite{bon76,blo79,cho87,sim01,sim03,uma05,mar05,sim07,sim09,ave13,sca13b,sca14}. 
More recently, it has also been used to investigate low-lying collective vibrations \cite{sim13b,sim13c,sca13b}, in particular to study their effects on fusion  \cite{was08,sim13b,sim13c,uma14}.

\begin{figure}
\centering
\includegraphics[width=5cm]{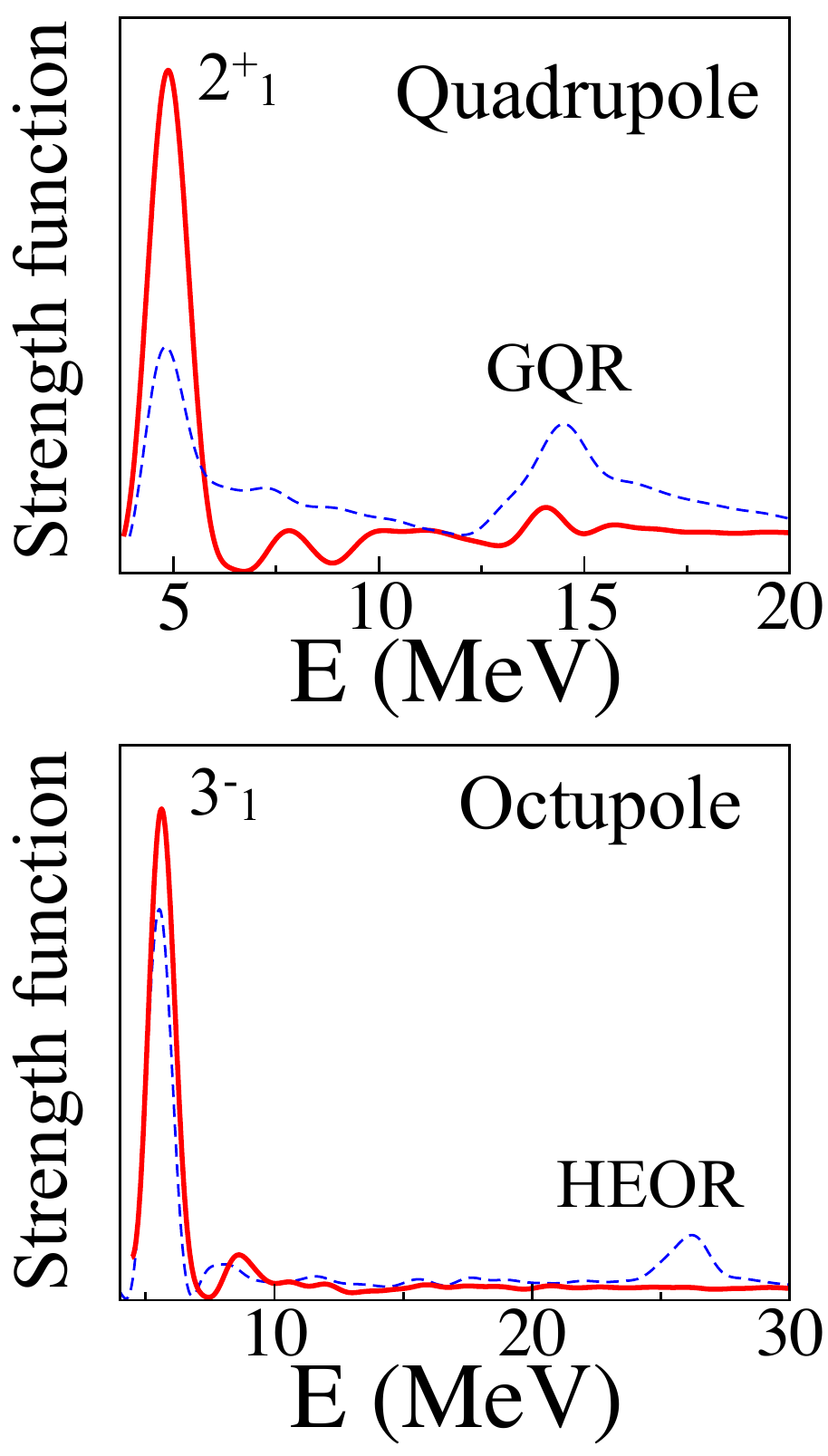}
\caption{Quadrupole (top) and octupole (bottom) strength functions in the $^{132}$Sn fission fragments (solid lines) compared with RPA calculations of vibrational modes build on the $^{132}$Sn ground state. Adapted from Ref.~\cite{sim14}.}
\label{fig-8}       
\end{figure}

As an illustration of application to study  vibration in fission fragments, the quadrupole and octupole moments computed in the fragments of $^{264}$Fm symmetric fission are shown in Fig.~\ref{fig-8} \cite{sim14}.
A comparison with  RPA calculations of vibrational modes built on the $^{132}$Sn ground state shows that the low-lying $2^+$ and $3^-$ are both excited in the fission fragments. 
However, the high energy modes, i.e., the giant quadrupole resonance (GQR) and the high-energy octupole resonance (HEOR), are not populated. 
Similar effects are observed in fusion where the dynamics is dominated by low-lying collective excitations \cite{das98}. 

\section{Conclusions}

The dynamics of the  fission process in  $^{258,264}$Fm has been investigated near scission at the mean-field level with the time-dependent Hartree-Fock approach including dynamical pairing correlations. 
It is shown that the spherical magic gaps in compact symmetric modes leading to tin fragments are formed well before scission, indicating a pre-formation of the fragments. 
Accounting for the time-dependence of the  pairing correlations, leading to an evolution of the occupation numbers, is crucial to allow the system to fission and to reduce the dependence with the initial condition.  
The results are used to interpret the experimental mass and kinetic energy distributions. 
It is shown that at least part of the internal excitation energy is stored into low-lying collective vibrations, while, as in fusion, the dynamics is less coupled to high energy modes. 

Quantum fluctuations beyond the independent particle/quasi-particle picture need to be incorporated in the future in order to reproduce the experimental distributions. 
The mass and charge distributions in the final fragments of a TDHF evolution can be calculated with a particle number projection technique \cite{sim10b}.
However, these fluctuations are underestimated at the mean-field level \cite{das79}. 
A possible approach is to incorporate beyond mean-field fluctuations with the TDRPA \cite{bal84} which has already been applied to heavy-ion collisions \cite{mar85,bon85,sim11}.
Applications to fission (neglecting pairing correlations) are promising \cite{sca15}. 
However, numerical solution of the TDRPA including pairing still has to be done. 

 \bibliography{biblio}
%
%
%
%

\end{document}